# RESIDUAL STRESS ANISOTROPY IN THIN-FILM LITHIUM NIOBATE FOR STRESS-MANAGED MEMS

*Byeongjin Kim, Ian Anderson, Tzu-Hsuan Hsu, Ziqian Yao, Mihir Chaudhari, Sinwoo Cho, and Ruochen Lu,*
The University of Texas at Austin, US

## ABSTRACT

In this work, we present the first experimental study of residual stress and post-release beam deflection in 128° Y-cut thin-film lithium niobate (TFLN) on Si, revealing pronounced stress anisotropy with in-plane orientation. Using optical profilometry with curvature fitting, we extract gradient stress ($\sigma_1$) to build orientation-resolved stress maps across multiple film thicknesses (100, 220, 460 nm). For films in the 220–460 nm range, we identify stress-free in-plane orientations near ~55° and 125°, enabling extremely flat suspended beams. In contrast, ultra-thin 100 nm films exhibit shifted stress-free orientations near ~20° and 160°. Leveraging these orientations, we demonstrate extremely long suspended beams up to 2 cm in length, 10 μm in width, and 460 nm in thickness without collapse. These results establish in-plane stress anisotropy and thickness selection in TFLN as a practical design lever for mechanically stable, scalable, and stress-managed microelectromechanical systems (MEMS).

## INTRODUCTION

Residual stress is a critical factor in determining the reliability of thin-film MEMS, radio frequency (RF) devices, and optical systems. Excessive residual stress can cause devices to collapse, break, or degrade performance during release (Fig. 1). Similar concerns have been widely reported in thin-film systems, where tensile residual stress can lead to film cracking [1]-[2]. Compressive residual stress can cause peeling, buckling, or blistering of the layer stack [3]-[4]. Residual-stress distributions are also known to affect adhesion, fracture toughness of thin films [5], and shift the resonant behavior of microelectromechanical and nanoelectromechanical systems (MEMS and NEMS) [6]. However, when properly controlled, stress can also be exploited as a useful tuning mechanism. Engineered stress has been shown to enhance electrical conductivity [7], modify the dielectric permittivity [8], increase piezoelectricity [9], tailor magnetic anisotropy [10], and even improve acoustic resonator performance [11]. These examples highlight that residual stress is not merely a byproduct of fabrication but a design parameter that can be intentionally utilized to optimize device performance.

In lithium niobate (LN), the strong elastic anisotropy and piezoelectric coupling make device performance sensitive to residual stress and its in-plane distribution. More recently, suspended devices built in transferred thin-film lithium niobate (TFLN) have been widely adopted for RF resonators [12], filters [13], photonics [14], and optomechanical applications [15].

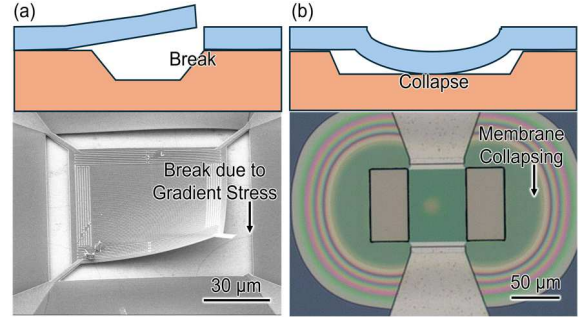

**Fig. 1**: Schematic and SEM/optical images of thin-film piezoelectric MEMS illustrating stress-induced failure modes: (a) membrane break due to stress gradient and (b) post-release membrane collapsing.

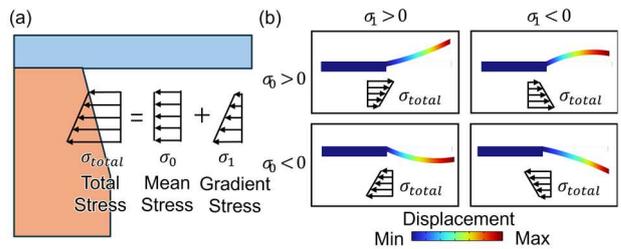

**Fig. 2**: (a) Decomposition of residual stress in a micromachined TFLN cantilever into mean $\sigma_0$ and gradient stress $\sigma_1$. (b) Four FEA-predicted post-release beam shapes produced by these stresses; colors denote displacement.

Residual stress in LN has been previously investigated in the context of bulk and thin-film acoustic devices. Thermal cycling in bulk SAW devices has been shown to induce significant residual stress that impacts reliability and crack formation [16]. Additionally, the influence of residual stress on acoustic wave propagation in piezoelectric LN layers has been investigated, revealing changes in wave slowness and coupling behavior [17]. Residual stress has also been observed in epitaxial LN films, where substrate clamping influences out-of-plane thermal expansion and phase stability [18]. However, to date, no report has addressed the thickness-dependent and in-plane orientation-dependent residual stress distribution in transferred TFLN. The lack of systematic stress characterization limits scalability and yield. This is especially important recently, as acoustic devices in TFLN are now operating at higher frequencies, necessitating the study of thinner films [19]-[20], which are more susceptible to the impact of stress.

In this work, we experimentally map the in-plane, thickness-dependent residual stress in 128° Y-cut TFLN and identify stress-free orientations near 55° and 125° for

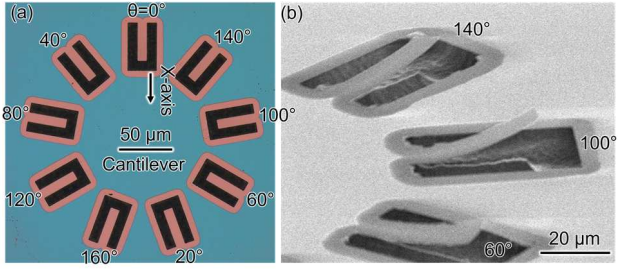

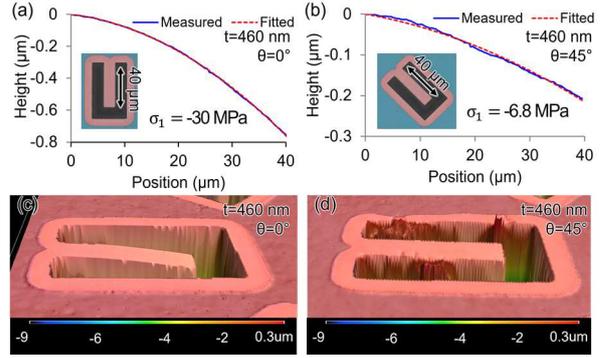

**Fig. 3**: (a) Optical image of an in-plane orientation array of TFLN cantilevers with θ varied from 0° to 160° in 20° increments. (b) SEM image of released cantilevers, highlighting orientation-dependent post-release beam curvature.

Figure 4: Optical profilometry of released TFLN cantilevers: (a–b) measured and fitted out-of-plane beam profiles with extracted gradient ($\sigma_1$) stress at θ = 0° and 45°, respectively; (c–d) corresponding 3D height maps of the same devices (colors indicate surface height).

220–460 nm films, which enable extremely flat suspended beams; further, we show that in ultra-thin 100 nm films, these stress-free orientations shift to ~20° and ~160°. Using these insights, we demonstrate centimeter-scale suspension (2 cm length, 10 μm width, 460 nm thickness) at θ = 132.5° without collapse, thereby establishing guidelines for in-plane orientation and thickness as a practical design strategy for scalable, mechanically stable TFLN MEMS.

## DESIGN AND SIMULATION

A uniaxial residual stress field in a thin film can be approximated by a first-order expansion through the film thickness [21]:

$$\sigma_{total} = \sigma_0 + \sigma_1 \left(\frac{y}{h/2}\right) \quad (1)$$

where $h$ is the film thickness and $y \in (-h/2, h/2)$ is the thickness coordinate. With this truncation, the stress is separated into a uniform (mean) component $\sigma_0$ and a linear gradient component $\sigma_1$. $\sigma_0$ is symmetric about the mid-plane, while $\sigma_1$ represents a stress gradient in the thickness direction. In typical thin film on substrate setups, $\sigma_0$ can be associated with the coefficient of thermal expansion (CTE) mismatch between the film and the substrate [22], whereas $\sigma_1$ represents the depth-dependent effects such as ion-implantation or bonding damage, or differential relaxation during chemical-mechanical polishing (CMP) [23].

To experimentally distinguish these two contributions, we adopt the cantilever-based methodology in [18], which decomposes the out-of-plane deflection profile of released beams into $\sigma_0$ and $\sigma_1$ components (Fig. 2a). Finite element analysis (FEA) (Fig. 2b) further illustrates how positive and negative $\sigma_0$ and $\sigma_1$ lead to distinct curvature and rotation signatures, providing intuitive guidance for identifying the dominant stress component from beam shape. Based on this framework, the corresponding stress components can be extracted quantitatively for cantilevers, following the stress–curvature calibration in [24].

$$\sigma_1 = \frac{Eh}{2R} \quad (2)$$

where $E$ is the Young's modulus and $R$ is the radius of curvature. The angular rotation θ is extracted from the beam profile. Because LN is anisotropic, $E$ must be taken as a directional modulus. After rotating the stiffness tensor into the 128° Y-cut frame [25], the effective modulus is obtained from the rotated compliance element $S'_{11}$, which corresponds to strain under uniaxial loading along the beam direction, i.e., $E(\theta) = 1/S'_{11}(\theta)$ [26]. Using this substitution, equation (2) remains valid with $E$ replaced by $1/S'_{11}$, enabling direct evaluation of the stress components for any beam orientation in 128° Y-cut TFLN. We note that the magnitude of $\sigma_1$ can vary with in-plane orientation in 128° Y-cut TFLN. This variation does not arise from elastic anisotropy, but rather from the underlying through-thickness residual stress gradient. We attribute this gradient to the anisotropic thermal expansion of LN, which causes different degrees of stress relaxation between the top and bottom of the film during cooldown. As a result, $\sigma_1$ reflects the through-thickness residual stress profile rather than elastic effects.

In isotropic thin films, both $\sigma_0$ and $\sigma_1$ can be reliably extracted from released beam profiles using curvature–rotation relationships [21]. However, this decomposition does not directly carry over to anisotropic TFLN. Unlike isotropic SiO₂, the total rotation angle θ in LN is strongly influenced by anisotropic bending stiffness and the localization of stress near the release junction, causing θ to vary with beam width and length. As a result, $\sigma_0$ becomes highly geometry-dependent and is not robust to be extracted reproducibly in 128° Y-cut TFLN with the current beam curvature-based method.

In contrast, $\sigma_1$ is determined primarily by the normalized curvature $h/R$, and is insensitive to both beam width and length, making it a more reliable metric across film thicknesses and orientations. We further validated this approach using FEA, which confirmed strong agreement between measured and predicted curvature. Therefore, in this work, we focus on $\sigma_1$, the gradient stress, as the primary measure of in-plane residual stress anisotropy in suspended 128° Y-cut TFLN.

## FABRICATION AND MEASUREMENT

Micromachined cantilever arrays were fabricated from 128 ° Y-cut TFLN on Si with thicknesses of 100 nm, 220 nm, and 460 nm, provided by NGK Insulators. While each set comprised beams spanning $\theta = 0°–160°$ in 20° increments (Fig. 3), four such sets were rotated by 5°, yielding a combined array covering $\theta = 0°–175°$ in 5° increments (36 groups in total for each thickness). The etch windows were patterned using low-temperature ICP-RIE to minimize the generation of thermal residual stress, followed by the release of the suspended structures using $XeF_2$. Beam deflections were measured using a laser confocal optical profilometer (Fig. 4a–b) and fitted with a quadratic model [21], while 3D height maps validated the curvature (Fig. 4c–d). Each orientation group was measured at three distinct chip locations to ensure reproducibility. Together, this platform enabled in-plane orientation- and thickness-resolved stress extraction, providing a mapping of in-plane stress anisotropy in TFLN.

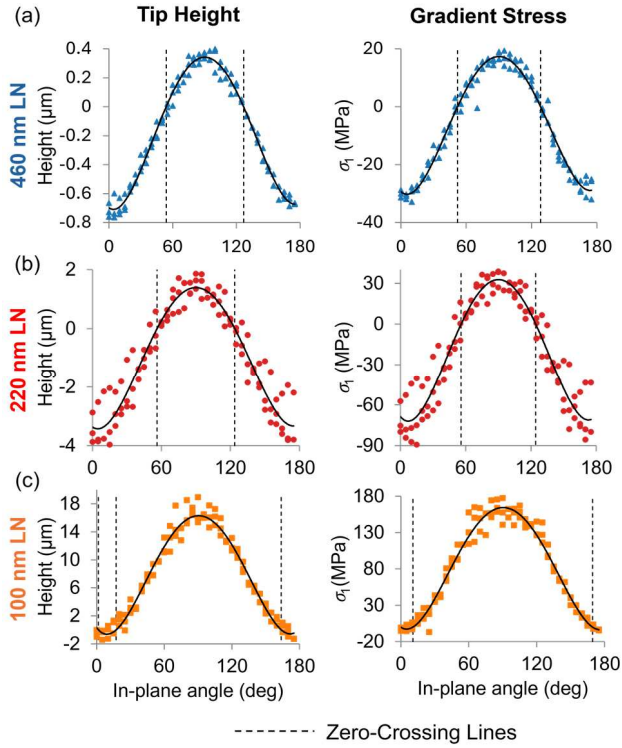

*Figure 5: Beam tip height and gradient stress ($\sigma_1$) versus in-plane angle in the TFLN of different film thicknesses: (a) 460 nm, (b) 220 nm, and (c) 100 nm. Points are measurements and lines are fits. To our knowledge, this highlights the first systematic study of the in-plane orientation dependence of stress in TFLN.*

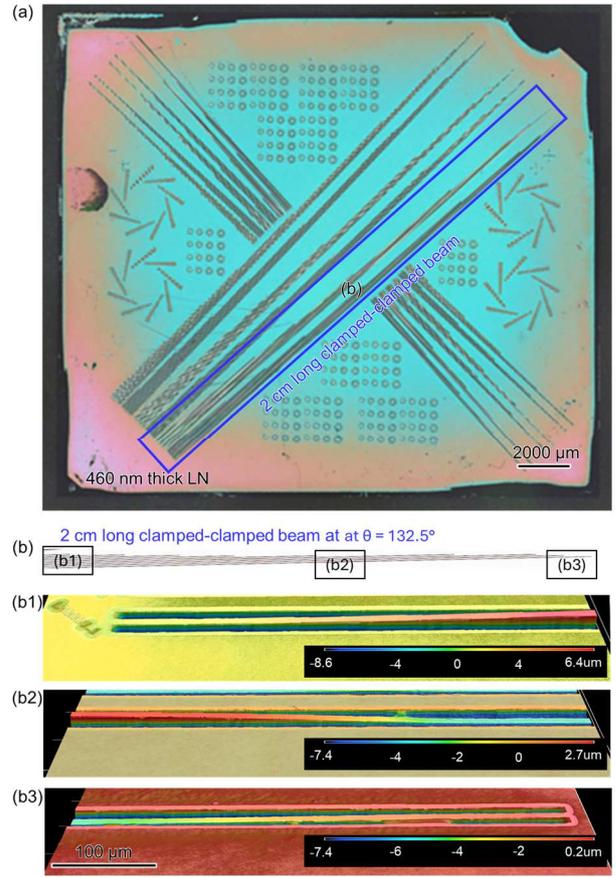

*Figure 6: (a) Optical micrograph of full sample with (b) 2-cm-long clamped–clamped beam. (b1) Left end, (b2) center, and (b3) right end of the 2-cm beam at $\theta = 132.5°$ in 460 nm thick TFLN.*

Figure 5 summarizes the extracted beam tip heights, along with the $\sigma_1$ across different thicknesses. For both 460 nm and 220 nm films, $\sigma_1$ crosses zero near 55° and 125°, yielding stress-free orientations that produce exceptionally flat cantilevers. Importantly, the orientations at which the beams appear flat closely coincide (within ~5°) with the zero-crossings of $\sigma_1$, suggesting that the residual mean stress $\sigma_0$ plays a relatively minor role in determining the shape of the suspended beam. Thus, the $\sigma_1$ serve as the primary descriptor of in-plane residual stress in these thin film systems. Building on this finding, Fig. 6 demonstrates a 2-cm-long clamped–clamped beam at $\theta = 132.5°$ in a 460 nm TFLN that remained fully suspended without collapsing, representing an unprecedented survivability length for TFLN beams (though slightly offset from the optimal 125° orientation). In addition, Fig. 5 reveals that in ultra-thin 100 nm films, the stress-free orientations shift to 20° and 160°. Notably, the cantilevers exhibit a larger upward curvature compared to the thicker films, indicating a larger positive $\sigma_1$ (i.e., a more tensile upper surface). This highlights a stronger thickness-dependent stress asymmetry in ultra-thin TFLN. These results collectively validate that careful orientation and thickness selection can mitigate stress-induced deformation, enabling the scalable fabrication of suspended MEMS in TFLN.

## CONCLUSION

In this work, we experimentally mapped the in-plane, thickness-dependent residual stress in 128°Y-cut TFLN on Si by extracting the gradient stress component $\sigma_1$ from the released cantilever curvature. We identified stress-free orientations near ~55° and ~125° for 220–460 nm films, which enable exceptionally flat suspended structures, and demonstrated a 2-cm-long, 10-µm-wide, 460-nm-thick beam at $\theta = 132.5°$ that remained fully suspended. In contrast, ultra-thin 100 nm films exhibited shifted stress-free orientations (~20° and ~160°) and consistently larger upward curvature, indicating a stronger thickness-dependent stress asymmetry. These findings establish in-plane orientation and film thickness selection as practical design parameters for mechanically robust, scalable suspended MEMS in TFLN. Looking ahead, applying crystallographic stress characterization techniques such as X-ray diffraction (XRD) or Raman spectroscopy will allow a more accurate evaluation of the $\sigma_0$, enabling a more complete understanding and control of residual stress across all thickness regimes.

## ACKNOWLEDGEMENTS

The work was supported by DARPA Chemistries and monoLayers for Anti-aging Kinematics (CLOAK) program. The authors would like to thank Dr. Sunil Bhave, Dr. Weileun Fang, and Dr. Chao-Lin Cheng for helpful discussions. Any opinions, findings, conclusions, or recommendations expressed in this material are those of the author(s) and do not necessarily reflect the views of the Defense Advanced Research Projects Agency (DARPA)

## CONTACT

*B. Kim: bk24672@my.utexas.edu